# Drying of a Microdroplet of Water Suspension of Nanoparticles: from Surface Aggregates to Microcrystal

**G. Derkachov,* K. Kolwas, D. Jakubczyk, M. Zientara, and M. Kolwas**

*Institute of Physics of the Polish Academy of Sciences, Al.Lotników 32/46, 02-668 Warsaw, Poland*

The method of formation of nanoparticle aggregates such as high-coverage spherical shells of microspheres or 3-D micro crystals grown in the geometry unaffected by a substrate is described. In the reported experiment, the evaporation of single levitated water droplet containing 200 nm diameter polystyrene spheres was studied. Successive stages of the drying process were discussed by analyzing the intensity of light elastically scattered by the evaporating droplet. The numerically simulated self-assembly coincides nicely with the observed morphologies resulting from transformation of a droplet of suspension into a solid microcrystal via kinetically driven self-assembly of nanostructures.

## 1. Introduction

Drying suspensions of nanoparticles can exhibit complex transitory structures during the process of evaporation of liquid. This phenomenon has recently been demonstrated for nanoparticles drying on a flat substrate.[1,2] The evaporation of liquid can drive the formation of nanoparticle monolayers[1] as well as the creation of aggregates of various morphologies. However, the presence of a substrate can dramatically influence the process of drying-mediated particles aggregation. Moreover, after evaporation of the liquid, migration of particles on the substrate usually continues (e.g., ref 3) and induces further self-assembly processes.

Much less is known about the drying mediated dynamics of particles ordering on curved surfaces in the absence of any substrate effects, which is unaffected by the presence of a substrate−liquid interface. The ideal realization of such a system for studying would be a suspension of nanoparticles forming unsupported spherical droplets.

In the present paper, we studied a single, unsupported aqueous droplet suspension of nanospheres levitating in electrodynamic quadrupole trap.[4,5] The evaporation-driven aggregation of nanostructures of nanospheres takes place in the spherical symmetry due to the surface tension of the liquid droplet at the conditions of the controlled thermodynamic parameters of the environment. The evaporation-driven aggregation of nanoparticles gives rise to the various transient surface and volume morphologies including dense packed spherical shells of nanospheres of long-range ordering. The observed process is terminated by formation of a solid microcrystal made of assembled nanospheres.

Under conditions reported in this experiment, the initial distribution of nanospheres is homogeneous in the droplet volume. As an effect of water evaporation, we expect the following scenario: Inclusions from the evaporated volume are gathered at the droplet surface. In the presence of attractive particle−interface interactions, nanoparticle surface islands nucleate and grow. The evaporation-driven aggregation of surface particles that we observe becomes more efficient than the self-assemblies aggregation inside the droplet volume. At this stage, the process is similar to the nanoparticles island formation studied in two-dimensional geometry reported in ref 1. Further evaporation and diminishing of the droplet surface leads to island rearrangement and to formation of regular, quasi-stable spherical crystalline shells of nanoparticles. These shells eventually collapse under the stress and form the solid, although still wet, crystal of nanospheres that rearrange toward the minimum energy configuration.

The reported experiment was performed on single drying water droplets containing spheres of standardized size. We studied the successive stages of the drying process by observing the light elastically scattered by the droplet. The changes in the intensity and polarization of the scattered light carry information about the dynamics of kinetically driven processes occurring at its surface and in its volume during evaporation. The optical characterization (section 3) of the droplet together with the numerical simulation of evaporation driven nanoaggregation (section 3.3) allowed us to propose a scenario of the transition from the initial aqueous droplet with homogeneously distributed inclusions, via the regular, quasi-stable spherical shell of nanoparticles filled with the colloid of nanoparticles, to the final microcrystallite.

## 2. Experimental Setup and Conditions

The details of the electrodynamic trap (Figure 1) construction and of the experimental setup can be found in.[5,6] The oscillating field in the quadrupole electrodynamic trap (e.g., ref 7) constitutes, for a charged (massive) particle, a pseudopotential. By the proper selection of driving parameters versus particle charge/mass ratio, a 3D (pseudo)potential minimum can be created in the center of the trap. The trap was mounted in the climatic chamber. The experiment was conducted at a temperature of 15 ± 0.3 °C, an atmospheric pressure of 1006 hPa, and a water vapor relative humidity of 94%. Special effort was made in order to avoid discharges between electrodes and thus enable operation in humid atmosphere.[5,6]

A droplet of homogeneous aqueous suspension was injected into the trap with a piezo-type droplet injector. The initial droplet radius was ∼15 $\mu$m, the initial volume concentration of polystyrene spheres in water was ∼1:1000, and their average diameter was 200 nm. The electric charge of the droplet was of the order of $5 \times 10^5$ elementary charges.[5] The charge was gained on exiting from the nozzle of the injector under the influenced of the electric field of the trap central electrode. Appropriate droplet injection timing versus the phase of AC voltage driving the electrodes allowed us to control the sign of the droplet charge, as well as, to certain extent, its magnitude.

* To whom correspondence should be addressed. E-mail: derkaczg@ifpan.edu.pl.

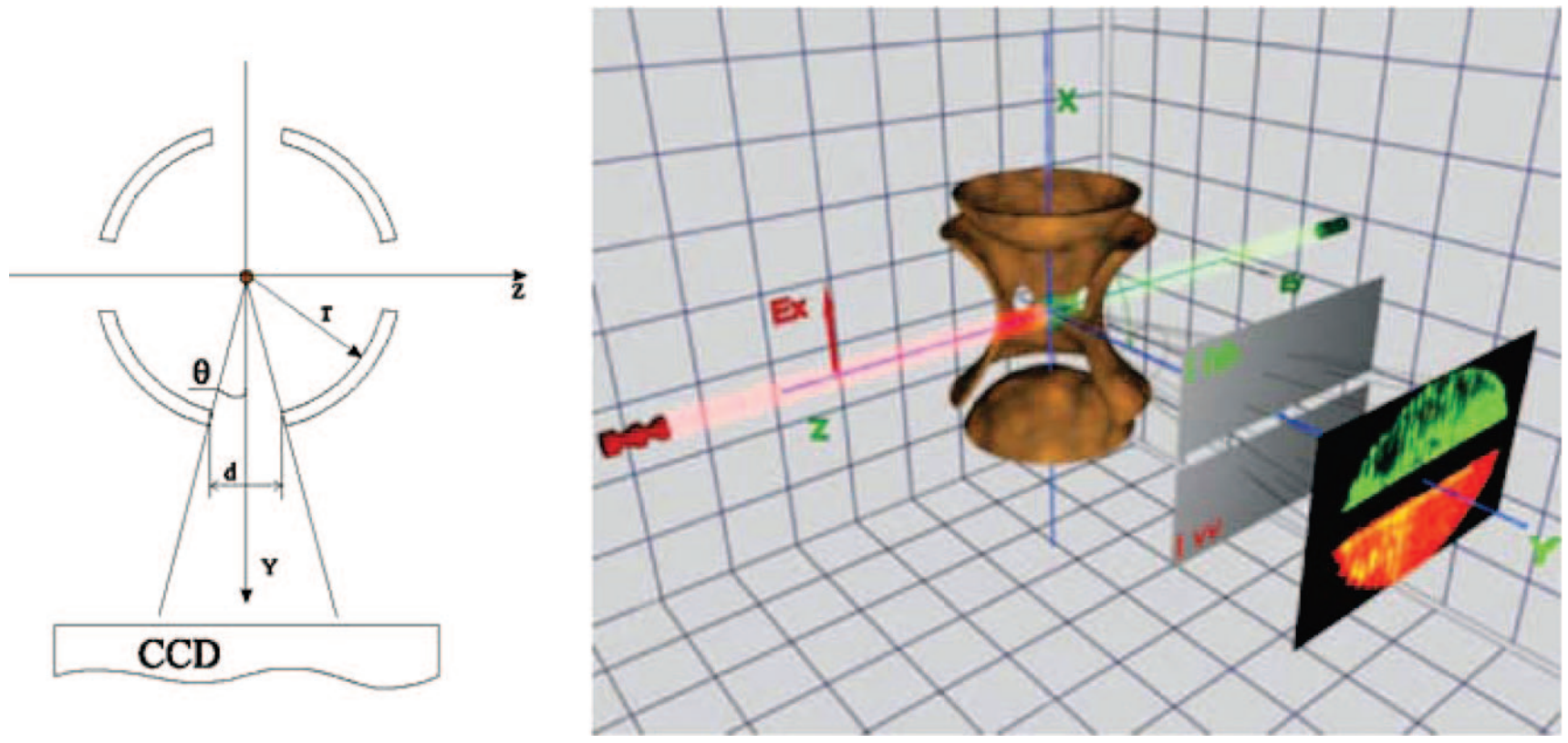

**Figure 1.** Scheme of the experimental setup representing the electrodynamic trap and two counterpropagating laser beams of orthogonal polarizations. Green and red fringes/speckles in the detection channel arise from the coherent light scattering by the trapped droplet/crystalite into the solid angle of $\Delta\Theta \sim 0.1$ sr around the right angle.

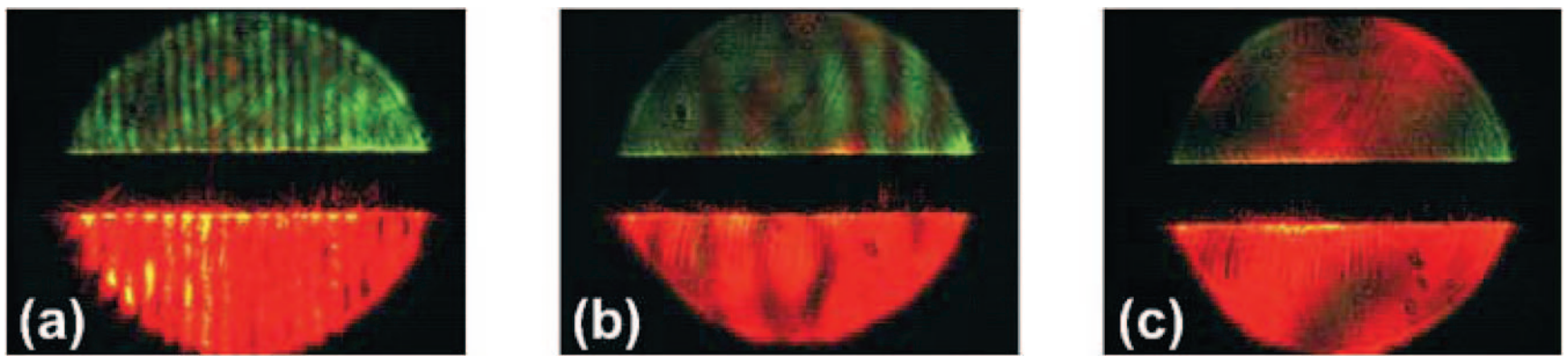

**Figure 2.** Examples of the interference patterns of light scattered by the trapped droplet/crystalite at different stages of water evaporation.

Two counterpropagating laser beams of orthogonal polarizations were used simultaneously (see Figure 1) for monitoring the droplet evolution. These were a He−Ne laser beam of 632 nm wavelength (red light, polarized perpendicularly to the observation plane ($s$ polarization)), ~18 mW CW, and a Nd:YAG laser beam of 532 nm wavelength (green light, polarization parallel to the observation plane ($p$ polarization)), ~25 mW CW. We recorded the spatial distribution of the intensity of light elastically scattered by the droplet/crystallite into the solid angle of $\Delta\Theta \simeq 0.1$ sr around the right angle in the observation plane, as shown in Figure 1. Two linear polarizers were used in the detection channel: first with the polarization direction parallel to the laser beam polarization (upper half of the channel) and the second perpendicular to it (the lower half of the channel). The spatial distribution of the intensity was registered with the 12-bit color digital CCD camera. It allowed us to separate spectrally the elastically scattered light and attribute it to appropriate incident beam polarization. We recorded temporal evolution of the intensity distributions of the scattered light of the same polarization as the polarization of the incident light (of a given color) ($I_{ss}(\Theta, t)$ and $I_{pp}(\Theta, t)$), and the cross-polarized (depolarized) light intensities ($I_{sp}(\Theta, t)$ and $I_{ps}(\Theta, t)$) correspondingly. The images were acquired by the camera at the frame rate of 40 frames/s and were converted into 16-bit color bitmaps with the algorithm ensuring high color fidelity. The intensities $I(\Theta, t)$ resulted from averaging the contributions for given azimuthal angle $\Theta$, over the available range of elevation angle.

The data acquisition was done with a PC in the form of a video. The examples of frames showing different stages of the evaporation process are presented in Figure 2.

The relatively regular interference pattern could be observed as long as the scattering particle was (at least partially) liquid. The distribution of the interference fringes/speckles carries information on the size of the droplet. Furthermore, it turned out that as far as the incident light is scattered by a spherical liquid droplet with homogeneously distributed inclusions, the Mie scattering theory,[8] is, to a certain extent, applicable.

## 3. Drying Droplet Characteristics Resulting from Light Scattering Measurements

The temporal evolution of the droplet radius $a(t)$ provides essential data about the dynamics of the kinetically driven processes occurring during evaporation of water from a droplet of nanosphere suspensions. The analysis of the scattered light intensities $I_{ss}(\Theta, t)$ and $I_{pp}(\Theta, t)$ of the same polarization as the polarization of the incident light, and the cross-polarized (depolarized) light intensity $I_{sp}(\Theta, t)$ enabled studying the size characteristics and some structural changes experienced by the droplet of aqueous suspension of nanoparticles.

**3.1. Drying Droplet Size Evolution.** The droplet radius temporal dependence $a(t)$ (Figure 3) was derived from the analysis of the frequency of the spatial modulation of the scattered light intensity $I_{ss}(\Theta, t_i)$ corresponding to successive frames of the registered video (examples in Figure 2).[5,6]

The method of finding the droplet radius by scattered light spacial frequency analysis[10] consists in the simple relationship between the spatial frequency $\omega_s$ of $I_{ss}(\Theta)$ and the size parameter $X = 2\pi a/\lambda$, where $\lambda$ is the wavelength of the scattered light field. This relationship was derived with a little help from Mie scattering theory[8] but can be perceived as partially phenomenological. In the range of size parameters $0 < X < 300$, this relationship can be expressed as $\omega_s \cong kX$ with $k = 4.83 \times 10^{-3}$. $\omega_s$ can be found by applying fast Fourier transform (FFT) to $I_{ss}(\Theta)$. This relationship is practically independent of the (effective) refractive index of the droplet. Therefore, the method is also applicable to the case of drying droplet with inclusions described by the effective index of refraction which changes inevitably during the water evaporation. We checked accuracy of the approach for the refractive indices in the range $1.33 < n < 2.2$. The method of spacial frequency analysis provided trustworthy information on the dynamics of the effective size evolution as long as the distinct interference pattern was visible (Figure 3). As long as the droplet can be treated as a spherical, globally homogeneous object, its radius can be found with the accuracy better than 8%. With the decreasing amount of water in the scattering structure, the distinct interference pattern was replaced by irregular speckles due to the strong gradients of optical properties of the scattering structure.

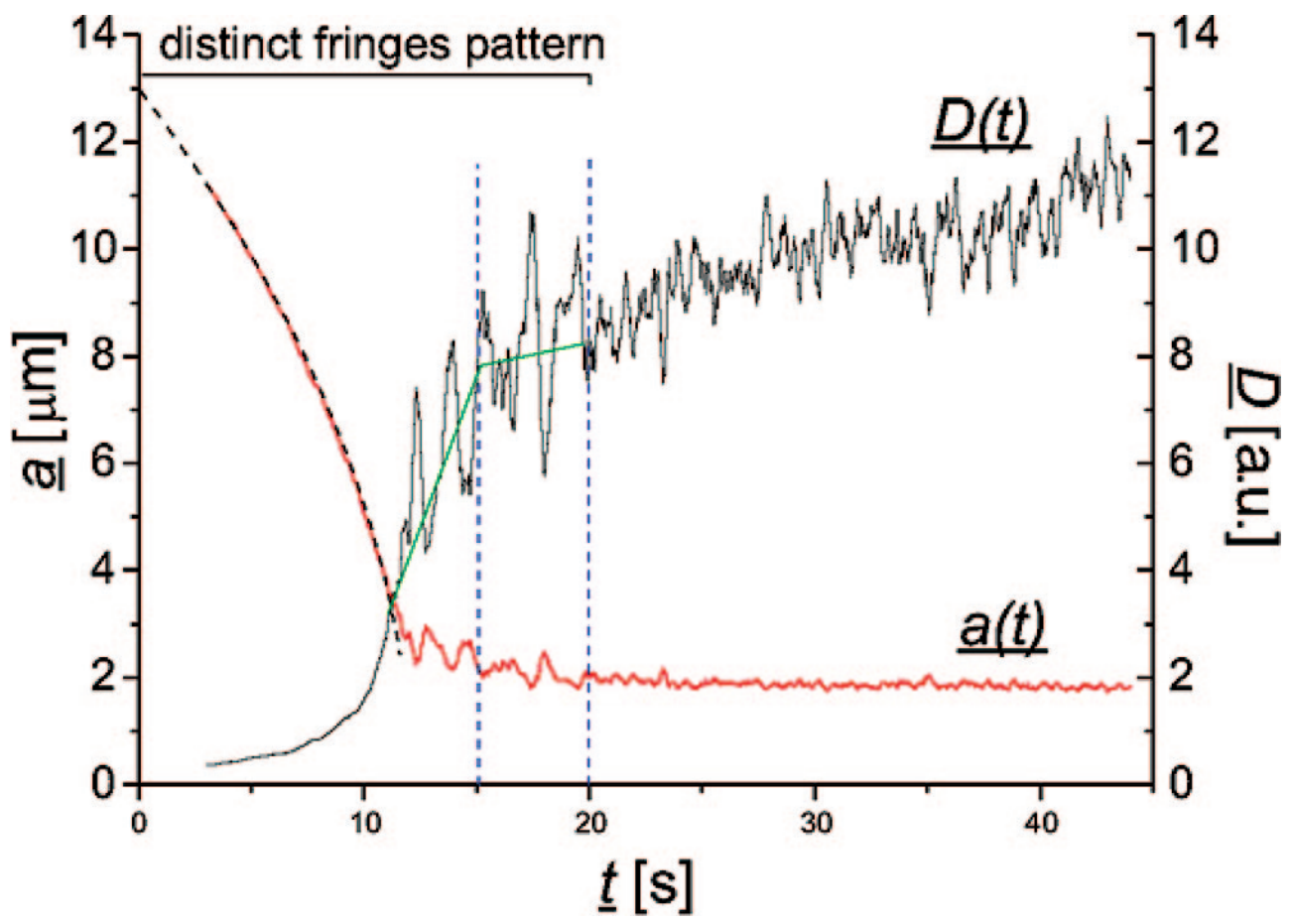

**Figure 3.** Left axis: the effective radius $a(t)$ for the drying droplet of aqueous nanospheres solution (red solid line) and for the radius temporal dependence of pure water (dotted black line). Right axis: Depolarization ability $D(t)$ (black solid line) used as a measure for the inhomogeneity of inclusions distribution.

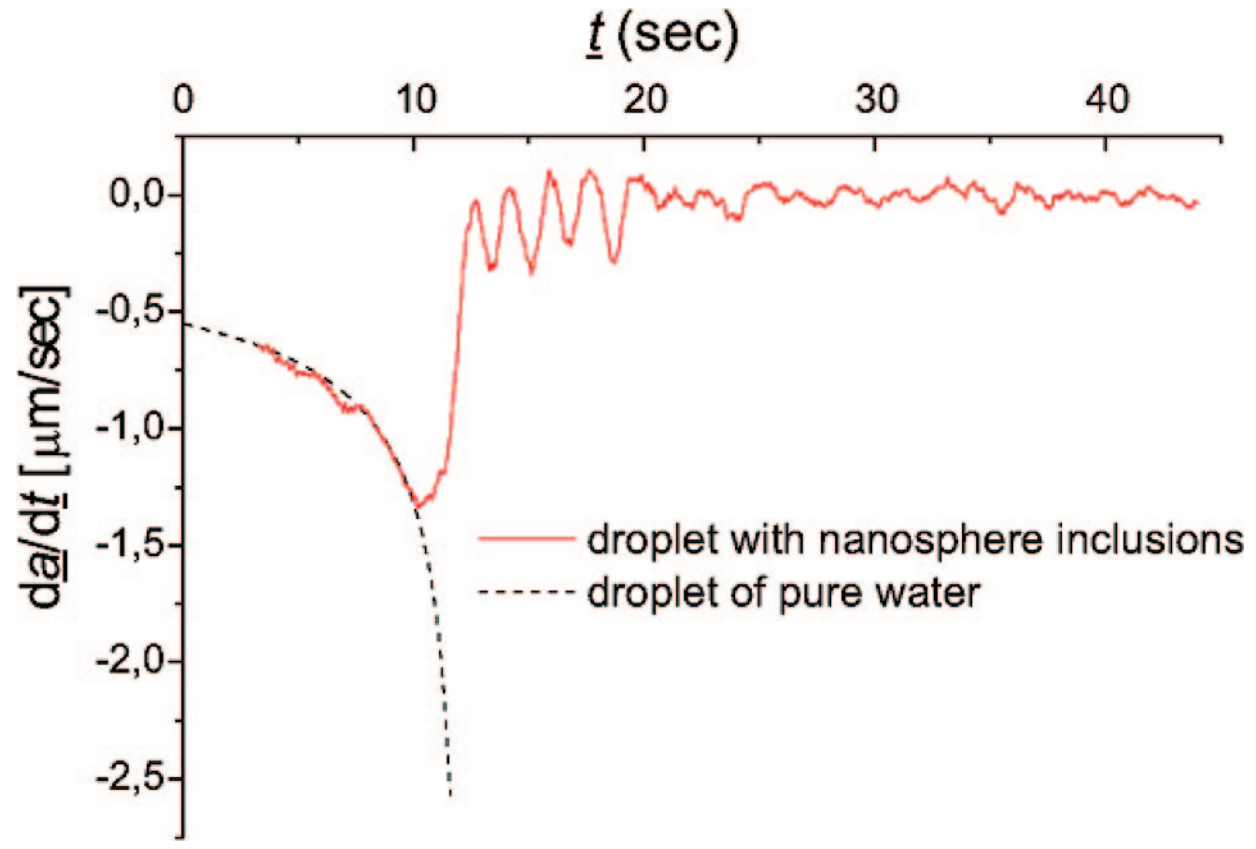

**Figure 4.** Comparison of the rates of radius changes of the drying droplet of aqueous nanospheres solution (solid red line) and of pure water (dotted line).

As the scattering object can rotate in the trap, the effective "optical" radius $a(t)$ shows significant fluctuations at some stages of the evolution. Such fluctuations result from the changes of scattering properties of the particle in the direction of signal detection rather then the instantaneous changes of the real size of the scattering object.

Solid red lines in Figures 3 and 4 correspond to the radius evolution of the drying water droplet with inclusions $a(t)$ (left axis) and the dotted line represents the radius evolution of a pure water droplet for the same thermodynamic conditions found from the model.[9,11]

The evaporation of pure water droplet is terminated by the instability associated with the Rayleigh limit (the Coulomb explosion) (see ref 12 and references therein), whereas for the contaminated droplet, the evolution can continue until the size becomes stabile.[9,11] As illustrated in figure 4, the rate of evaporation $|da/dt|$ can be dramatically slowed down by growing concentration of inclusions, as discussed in section 4 below.

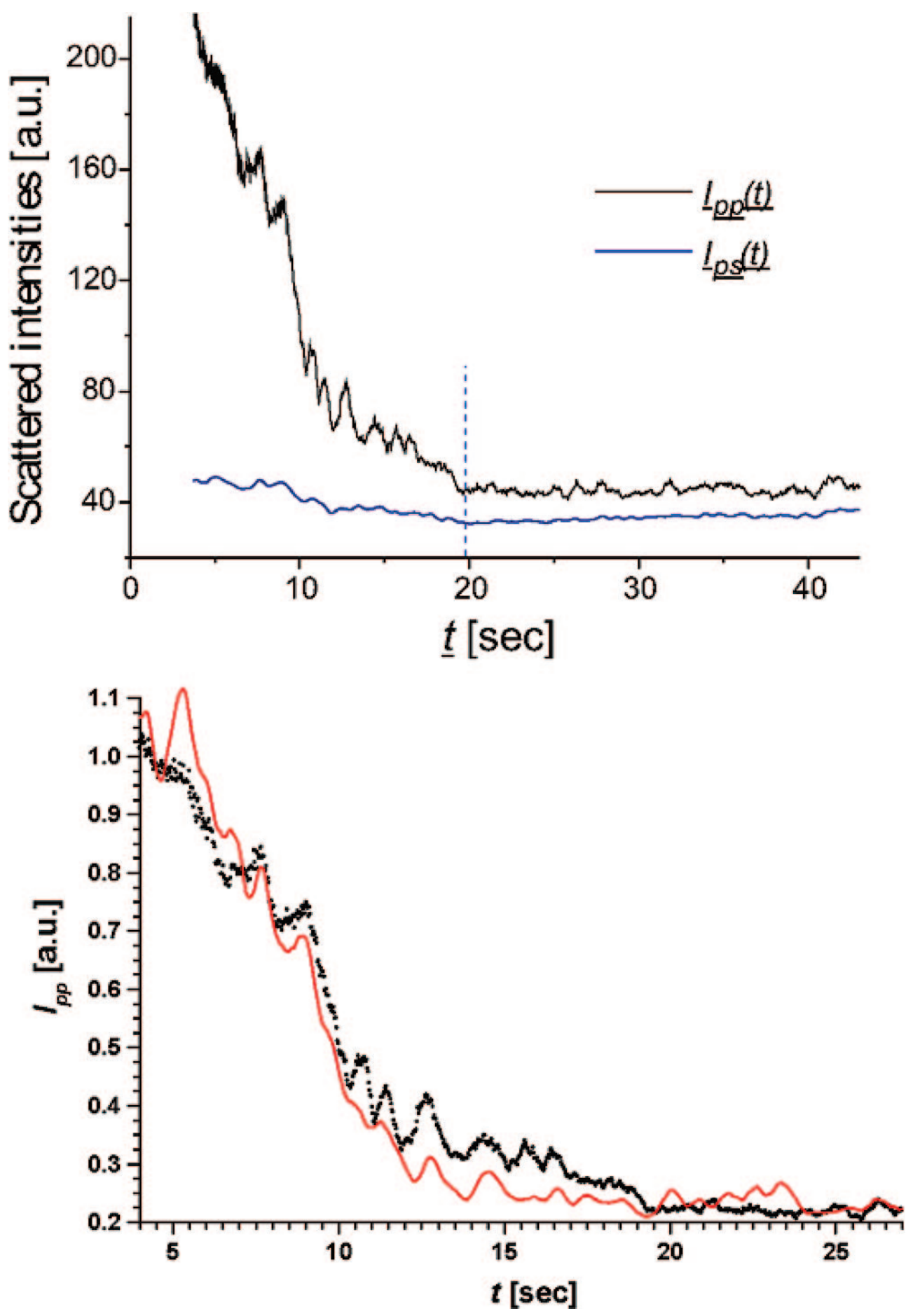

**Figure 5.** Top: The scattered light intensity $I_{pp}(t)$ of the same polarization as the polarization of the incident light (black line) and the cross-polarized (depolarized) light intensity $I_{ps}(t)$ (violet line). Bottom: Experimental (black dots) and smoothed (15-points adjacent average) theoretical (solid red line) $I_{pp}(t)$.

**3.2. Evolution of Concentration Inhomogeneity of Inclusions.** The total intensities $I_{pp}(t)$ and $I_{ps}(t)$ (as well as $I_{ss}(\Theta, t)$ and $I_{sp}(\Theta, t)$) of light scattered into the observation angle integrated over the observation angle $\Theta$ respectively, provide valuable information on drying droplet properties. $I_{ps}(t)$ is almost 2 orders of magnitude smaller than $I_{pp}(t)$ (Figure 5) Both signals follow the droplet size and composition (altered by the droplet size) changes.

The scattered cross-polarized signals $I_{ps}(t)$ are absent for homogeneous spheres: both in experiments with pure liquid droplets and in Mie theory. Our previous experiments with pure water droplets[5,11] excluded the possibility of depolarization by small deformations of the spherical droplet kept in the trap.[8,13,14] Therefore, the observed depolarized light $I_{ps}(t)$ is due to the departure of optical properties of the drying droplet with inclusions from the corresponding properties of a pure liquid droplet. The depolarized light intensity $I_{ps}(t)$ per unit surface of the sphere $D(t) = I_{ps}(t)/\ a^2(t)$ can serve as a measure for the inhomogeneity of the inclusions distribution. $D(t)$ presented in Figure 3 illustrates the depolarization ability evolution of the drying suspension droplet, caused by the volume and surface inclusions distribution (inhomogeneity) changes. However, the nanosphere aggregates formed at the drying droplet surface are expected to have higher impact upon the depolarization signal than those in the volume. As we expect, shrinking of the droplet surface introduces significant density fluctuations of inclusions near the surface and leads to significant gradients of refractive index. Such situation persists as long as the scattering object can be treated as a droplet rather than a wet assembly of nanospheres.

Up to the 20th second of the droplet evolution, the $I_{pp}(t)$ dependence could be relatively (and surprisingly) well reproduced by Mie scattering theory into which the effective index of refraction was introduced: $n_{eff}(a) = f(a)n_{ns} + (1 - f(a))n_v$ (see, e.g., ref 15; Figure 5, bottom). $f(a)$ is the volume fraction of nanospheres in the liquid, changing with the progress of evaporation, $n_w = 1.33$ is the index of refraction of water, and $n_{ns} = 1.5$ is the index of refraction of the nanosphere material. It must be underlined that $a(t)$ was found from $I_{ss}(\Theta, t)$ dependence (another polarization) with the method described in section 3.1 (and not with the Mie theory). The applicability of the Mie theory in this case is somewhat unexpected, since it has been formulated for homogeneous (and isotropic and linear) medium. The transformation of the droplet of suspension into the wet assembly of nanospheres possibly takes place at ~20th second. It is followed by a slight continuous increase of both $I_{pp}(t)$ and $I_{ps}(t)$ intensities (Figure 3) in spite of a simultaneous decrease of the effective radius $a(t)$ (Figure 3). At this stage $I_{pp}(t)$ can not be reproduced by Mie scattering theory using the effective refraction index $n_{eff}$ as a parameter. The effective index of refraction of the microcrystallite depends on additional parameters such as the close packing fraction of the nanospheres, the content of water inside the crystallite and the wetting ability the nanospheres. Therefore its analytic form would have to be more complicated.

**3.3. Simulation of Topographical Nanostructure Patterning.** The dynamics of evaporation-driven nanostructures assembly was numerically simulated (in MATLAB and SIMULINK; for our programs see ref 16) for the water droplet containing 500 polystyrene nanosphere of 200 nm diameter. The dynamics of nanoparticle assembly and the shapes of intermediate and final structures are determined by the relative timescales of evaporation, and nanoparticle mobility in the medium. The random Brownian movements of nanospheres in liquid leads to their sticking. A tendency of a system to progress to a state in which the entropy is maximized can be perceived as entropic effective forces acting between hard spheres. Entropic forces in colloidal suspensions are known to be responsible for the crystallization of hard spheres. In our simulation such forces are modeled with the Lenard-Jones potential. The movement of inclusions is described with the Newton equation with the flow resistance defined by the viscosity coefficient. As the nanosphere hits the liquid interface it is, in our model, additionally subjected to the surface adsorbing force, which is assumed to be proportional to the distance between the nanosphere center and the interface. The rate of water evaporation from the droplet was taken from the experiment (smoothed $a(t)$, Figure 3). A survey of snapshots of aggregate morphologies obtained is provided in Figure 6 A video version serving as a source for Figure 3 is available in ref 17.

Several regimes of evaporation-mediated nanoparticles assembly can be noticed. At early stages of water evaporation (Figure 6a), the droplet exhibits (mostly) liquid surface. The density of single nanospheres is relatively small and the density of small aggregates of nanospheres is even smaller both at the surface and in the volume. With the progress of evaporation, a shrinking liquid surface gathers inclusions from the outer, evaporated volume. As a result a formation of islands of nanospheres at the droplet surface takes place (Figure 6, panels b and c). For longer times, high-coverage patterns develop and tend toward large islands (Figure 6, panels b and c) that rearrange forming the regular, quasi-stable layer of nanoparticles (Figure 6, panels d and e). At this stage only the slight rearrangements of nanospheres at the surface can be noticed. Further progress of evaporation results in a collapse of the shell of nanoparticles and formation of the solid, although still wet crystal of assembled nanospheres (see Figure 6f). The nanospheres within the structure and at its surface can still rearrange as the crystal dries (Figure 6g).

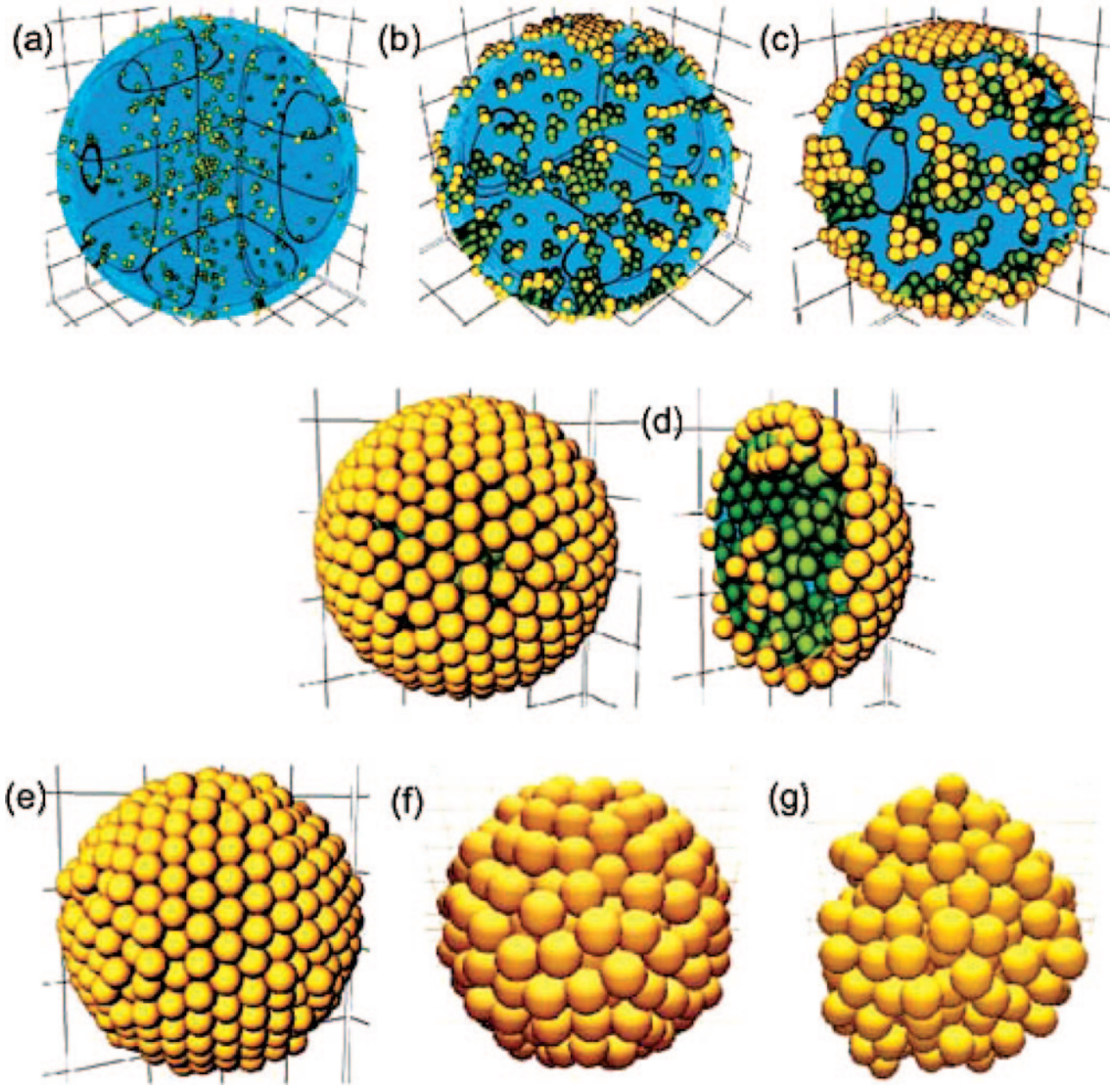

**Figure 6.** Survey of aggregate morphologies resulting from numerical simulations illustrating the drying mediated dynamics of particle ordering (the droplet/crystallite size was normalized).

The number of inclusions taken into consideration was by the order of magnitude lower than that in the reported experiment. It was due to the computing time saving. In spite of that limitation, and the simplicity of the particle movement model, all the characteristic stages of the droplet/microcrystallite dynamics seem to be demonstrated. We expect, that for substantially higher number of inclusions the formation of the cover shell could followed by successive collapses should repeat, as discussed in section 4.

## 4. Discussion of Evaporation Driven Processes

Our discussion of experimental results is based on the observation of several experimental events and involves both the optical characterization, in section 3, and the nanostructure patterning simulation, in section 3.3. The results suggest the existence of several regimes of evaporation-mediated nanoparticles assemblies, including the long-range ordering layer formation at the surface of the droplet as well as the final volume crystal of nanospheres.

During the first 10 s of water evaporation the fringes in the interferograms are regular (Figure 2a). The resulting $a(t)$ dependence (Figure 3) reflects the droplet size evolution. The mean rate of radius changes $|da/dt|$ follows the corresponding rate for the evaporating droplet of pure water (Figure 4). This stage of evolution can be illustrated with Figure 6a. However, the self-assembly process develops at the surface under increasing surface nanospheres concentration that results from decreasing volume of the droplet. A shrinking liquid surface gathers inclusions from the outer droplet mantle and drives the formation of islands of nanospheres. Increasing inhomogeneity of the nanospheres concentration results in slow increase of the depolarization ability $D(t)$, as shown in Figure 3.

Starting from the 10th second, the rates of radii changes for the droplet of pure water and for the droplet with inclusions become different, as illustrated in Figure 4. The nanosphere islands film gradually covers the decreasing surface of the droplet, which hinders the evaporation process. As a result, the rate of radius change $|da/dt|$ decreases, as illustrated in Figure 4. This stage of evolution can be illustrated with Figure 6, panels b and c. Inhomogeneities of the inclusions concentration at the surface cause further increase of the mean depolarization ability $D(t)$ (Figure 3). The interference fringes are still visible (Figure 2b), so the walking average of $a(t)$ is still a measure of the radius temporal dependence in spite of large $a(t)$ fluctuations caused by surface and volume inhomogeneities. As the droplet in the trap can rotate, the effective radius $a(t)$, exhibits also significant fluctuations due to the changes of directional optical properties of the droplet.

Between the 15th and 20th second of the evolution, the mean level of the depolarization ability $D(t)$ increases at smaller rate and is accompanied by only a small decrease of the mean $a(t)$ slope. As our simulation suggests (Figure 6, panels d and e), a dense, closely packed nanospheres film covering the drying surface is formed. Only slight rearrangements of nanospheres of the cover can be noticed. The quasi-stable shell of nanospheres surrounds a colloid of nanoparticles. As the evaporation of water proceeds, it might be possible, that after a shell collapses, and a new shell is built. Although there is no direct experimental proof of such scenario, the observations are not in contradiction.

At the 20th second, a change in optical properties of the scattering structure can be noticed. The regular fringes of the scatterograms are smeared out (see Figure 2c). We interpret it as a final collapse of the surface shell of nanoparticles the formation of a solid, although still wet crystal of nanospheres (see Figure 6f). At this stage, the FFT method of finding the droplet size (section 3.1) becomes inadequate and $a(t)$ presented in Figure 3 can not be treated as particle radius but only as some optical measure of evolution of the drying particle. The formation of wet crystal assembly modifies the optical properties of the evolving object and results in an increase of $I_{pp}(t)$ and $I_{ps}(t)$ intensities, (and so of $D(t)$), though the size of the crystal possibly decreases as the nanospheres can still rearrange (Figure 6g) as it dries.

## 5. Conclusions

The reported process of formation of nanoparticle aggregates seems to offer some new possibilities of producing aggregates such as high-coverage spherical shells of nano- or microspheres[18] or microcrystals of desired composition and size. The spherical symmetry imposed by the surface tension of an unsupported liquid droplet gives unique opportunity for producing microcrystals of long-range ordering in the geometry unaffected by the flatness of a substrate present in many other aggregation techniques. The method is applicable to various materials. The inclusions can also be poly dispersive. The liquid must only exhibit attractive particle interface interaction. The dynamics of the evaporation-driven aggregation processes can be controlled by the thermodynamic parameters of the environment, the viscosity and evaporation coefficient of the liquid, the size, initial concentration of nanospheres and their composition as well as the surface adsorption ability. The evaporation-driven aggregation of nanoparticles gives rise to the various transient surface and volume morphologies. Different choices of liquid, nanoparticle size and composition as well as the evaporation rate, etc. can result in various morphologies of the structures.

We expect, that the reported method of nanostructure assembly could be of technological value. In particular, the closely packed spherical crystalline shells of nanoparticles, could be stabilized, e.g., by coating or by addition of a binder to the suspension. The formed microcrystallite could also be softly deposited on a substrate in a controlled manner and serve as a building block in construction of some metamaterials in diverse applications from biomedical[19] through materials science[20]to food processing.[21]

**Acknowledgment.** This work was supported by Polish Ministry of Education and Science under Grant No. 1 P03B 117 29.